\documentclass[12pt]{iopart}
\usepackage{rotating}
\usepackage{graphicx}
\usepackage{latexsym}
\usepackage{iopams}
\usepackage{bm}

\newcommand{\RGf}{\ensuremath{\mathcal{R}}}
\newcommand{\PGf}{\ensuremath{\mathcal{P}}}
\newcommand{\ave}[1]{\mbox{$\langle #1 \rangle$}}

\begin{document}

\letter{Perimeter Generating Functions For The Mean-Squared Radius Of Gyration Of Convex Polygons}

\author{Iwan Jensen}
\address{ARC Centre of Excellence for Mathematics and Statistics of Complex Systems, 
Department of Mathematics and Statistics, 
The University of Melbourne, Victoria 3010, Australia}

\date{\today}

\ead{I.Jensen@ms.unimelb.edu.au}

\begin{abstract}
We have derived long series expansions for the perimeter generating functions of the 
radius of gyration of various polygons with a convexity constraint. Using the series 
we numerically find simple (algebraic) exact solutions for the generating functions. 
In all cases the size exponent $\nu=1$.
\end{abstract}

\submitto{\JPA}

\pacs{05.50.+q,05.70.Jk,02.10.Ox}

\maketitle

\section{Introduction}

A well-known long standing problem in combinatorics and statistical mechanics is
to find the generating function for self-avoiding polygons (or walks) on a two-dimensional
lattice. The models are of tremendous inherent interest as well as serving as simple models of 
polymers and vesicles \cite{MSbook,HughesV1,RensburgBook}. Despite strenuous effort over the 
past 50 years or so this problem has not been solved on any regular two dimensional lattice.
However, there are many simplifications of this problem that are solvable \cite{BM96a}, 
but all the simpler models impose an effective directedness or other constraint that reduces 
the problem, in essence, to a one-dimensional problem. 

\begin{figure}
\begin{center}
\includegraphics{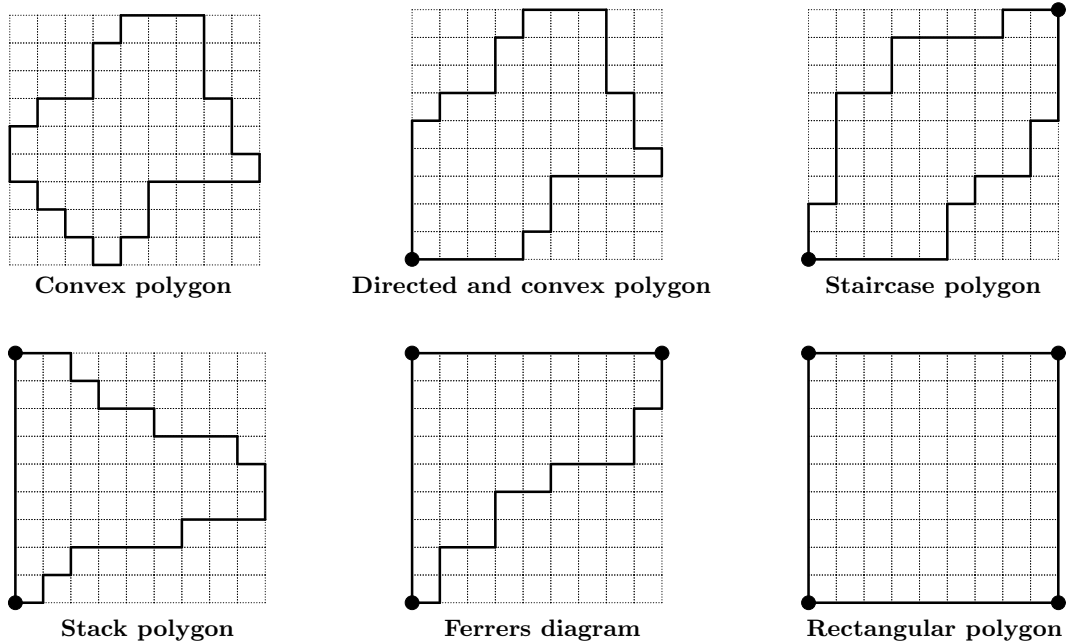}
\end{center}
\caption{\label{fig:poly} 
Examples of the types of convex polygons we consider in this paper.
}
\end{figure}

One particular class of exactly solved polygon models are those with a convexity constraint 
(see figure~\ref{fig:poly}). On the square lattice a polygon is said to be {\em convex} if it 
is convex with respect to both vertical and horizontal  lines, i.e.,  any vertical line will 
intersect the polygon at zero or two horizontal edges while similarly any horizontal line will 
intersect the polygon at zero or two vertical edges. Alternatively a convex polygon is a SAP 
of a length equal to the perimeter of its minimal bounding rectangle. If we further demand 
that the polygon must include the vertices in some of the corners of the minimal bounding 
rectangle we can define a further five polygon models as illustrated in  figure~\ref{fig:poly}. 
The full perimeter and area generating functions are known for all these models \cite{BM96a}. 
Also of great interest is the mean-square radius of gyration, $\ave{R^2}_n$, which measures 
the typical size of a polygon with perimeter $n$. In this paper we report on work leading 
to conjectured exact solutions for the generating functions associated with the mean-square 
radius of gyration for the class of convex polygons.

An {\em $n$-step self-avoiding walk} $\bm{\omega}$  is a sequence of {\em distinct} vertices 
$\omega_0, \omega_1,\ldots , \omega_n$ such that each vertex is a nearest neighbour of it 
predecessor. SAWs are considered distinct up to translations of the starting point $\omega_0$.
A self-avoiding polygon of length $n$ is an  $n-1$-step SAW such that $\omega_0$ and $\omega_{n-1}$ 
are nearest neighbours and a closed loop can be formed by inserting a single additional step 
joining the two end-points. We shall use the symbol $\bm{\Omega}_n$ to mean the set of all SAPs 
of perimeter length $n$. Generally SAPs are considered distinct up to a translation, so if there 
are $p_n$ SAPs of length $n$ there are $2np_n$ walks (the factor of two arising since the walk 
can go in two directions). One expects in general that $p_n \sim A\mu^n n^{\alpha-3}$, where 
$\mu$ is the so-called connective constant while $\alpha$ is a critical exponent. In our cases 
$\mu$ and $\alpha$ are known from the exact solutions for the perimeter generating functions

\begin{equation}\label{eq:genfunc}
\PGf (x) = \sum_n p_{2n} x^n \sim A(x)(1-\mu^2 x)^{2-\alpha},
\end{equation}
where we took into account that polygons on the square lattice have even length.
The generating functions thus have a singularity at the critical point $x_c = 1/\mu^2$ 
with critical exponent $2-\alpha$. The function $A(x)$ is analytic at $x=x_c$.
Note that both $\mu$ and $\alpha$ are model dependent.

The mean-square radius of gyration of $n$-step polygons is defined by,

\begin{equation}
\ave{R^2}_n = \frac{1}{2n^2 p_n} \sum_{\bm{\Omega}_n} 
\sum_{i,j=0}^{n-1} (\omega_i - \omega_j)^2,
\end{equation}
\noindent
where we expect that $\ave{R^2}_n \sim Bn^{2\nu}$.
It is advantageous to look at the quantity $r_n = n^2p_n\ave{R^2}_n$, which is an integer,
and in particular we shall study the associated generating function
\begin{equation}
\RGf(x) = \sum_{n} r_{2n}x^n \sim B(x)(1-\mu^2 x)^{-(\alpha+2\nu)},
\end{equation}
where we again used that $r_n$ is non-zero only when $n$ is even.

The values for the critical exponents are known exactly, though non-rigorously,  for 
self-avoiding polygons due to the work by Nienhuis \cite{Nienhuis82a},  $\alpha = 1/2$ 
and $\nu =3/4$. As we shall demonstrate later, the exponent $\alpha$ takes on several
different values for the convex polygons studied in this paper, but the exponent $\nu=1$
in all cases.

In the next section we briefly describe the algorithm used to calculate $r_n$ and in the
following section we list the various perimeter generating functions. 

\section{Computer enumeration \label{sec:enum}}

The first terms in the series for the polygon generating function are calculated 
using transfer matrix techniques to count the number of polygons spanning rectangles $W+1$ 
edges wide and $L+1$ edges long. The transfer matrix technique  involves drawing a line 
through the rectangle intersecting a set of edges. For each configuration of occupied or 
empty edges along the intersection we maintain a (perimeter) generating function for 
partial polygons cutting the intersection in that particular pattern. Due to the convexity 
constraint a vertical line will intersect the polygon exactly twice. The upper edge of the 
convex polygon performs a directed walk taking steps to the right and up until it reaches 
the top of the rectangle where it turns and then performs a directed walk with steps to the 
right and down. Likewise the lower edge performs a directed walk with right and down steps 
until it hits the bottom of the rectangle where it turns and takes only right and up steps. 
A convex polygon is formed once the two walks meet.  In order to specify a configuration we 
just need to know the positions of the edges and whether or not the top and bottom of the 
rectangle has been touched. All the possible configurations can then be encoded by four 
$(W+1)\times (W+1)$-matrices, one matrix for each possibility of touched borders. As the 
vertical boundary line is moved one step forwards the matrices are updated to allow for 
all the legal  moves of the edge-walks (the walks must be directed as described above and 
never cross). The updating involves simple double sums over the indices. This approach was used 
by Guttmann and Enting \cite{GE88b} and is very efficient. However, in one iteration many 
steps can be inserted and this makes the calculation of the contributions to the radius of 
gyration somewhat cumbersome. We find it more convenient to use an algorithm in which the 
convex polygons in a given rectangle are enumerated by moving the intersection so as to add 
one vertex at a time. The method we used to enumerate convex polygons on the square lattice 
is a specialisation of the method originally devised by Enting \cite{IGE80e} for the enumeration 
of self-avoiding polygons. As noted earlier, convex polygons can be viewed as SAPs 
with a number of steps equal to the perimeter of the minimal bounding rectangle. So we 
could simply take our previous algorithm \cite{JG99,IJ00a}, which we generalised in order 
to calculate the radius of gyration, and only extract the terms counting convex polygons.   
Due to the convexity constraint we were able to simplify the algorithm somewhat and
make it more efficient. However, the algorithm is still quite similar to the SAP enumeration
algorithm so we won't describe it further. Suffice to say that the method for calculating
the radius of gyration coefficients $r_n$ has been described in \cite{IJ00a}.

Using this algorithm we quickly (a few hours of CPU time) calculated the radius of gyration
of the polygon models of figure~\ref{fig:poly} to length $n=110$, giving us 56 terms in the
half-perimeter series. The first few terms $p_n$ and $r_n$ are listed in table~\ref{tab:coeff}.
The full series for the generating functions studied in this paper 
can be obtained by sending a request to the author or
via the web at http://www.ms.unimelb.edu.au/\~{ }iwan/.

\begin{table}
\caption{\label{tab:coeff} The number of polygons $p_n$ and their mean-squared radius of
gyration $r_n= n^2p_n\ave{R^2}_n$.}
\scriptsize
\begin{tabular}{rrr|rr|rr}
\br
  &  \multicolumn{2}{c}{Convex polygons}
  &  \multicolumn{2}{c}{Directed and convex polygons}
  &  \multicolumn{2}{c}{Staircase polygons} \\
\mr
$n$ & $p_n$ & $r_n$  & $p_n$ & $r_n$  & $p_n$ & $r_n$ \\
\mr
4 & 1 & 8 & 1 & 8 & 1 & 8 \\ 
6 & 2 & 66 & 2 & 66 & 2 & 66 \\ 
8 & 7 & 600 & 6 & 522 & 5 & 444 \\ 
10 & 28 & 5164 & 20 & 3772 & 14 & 2710 \\ 
12 & 120 & 41768 & 70 & 25138 & 42 & 15512 \\ 
14 & 528 & 317584 & 252 & 157212 & 132 & 84756 \\ 
16 & 2344 & 2280792 & 924 & 935140 & 429 & 446952 \\ 
18 & 10416 & 15573120 & 3432 & 5343160 & 1430 & 2291718 \\ 
20 & 46160 & 101743312 & 12870 & 29541450 & 4862 & 11485760 \\ 
22 & 203680 & 639664960 & 48620 & 158920172 & 16796 & 56486716 \\ 
24 & 894312 & 3889101336 & 184756 & 835390460 & 58786 & 273405288 \\ 
26 & 3907056 & 22961959168 & 705432 & 4305416136 & 208012 & 1305401916 \\ 
28 & 16986352 & 132118984560 & 2704156 & 21812985652 & 742900 & 6159651344 \\ 
30 & 73512288 & 743046249664 & 10400600 & 108875244952 & 2674440 & 28766573800 \\ 
32 & 316786960 & 4095077270128 & 40116600 & 536326527048 & 9694845 & 133128274320 \\ 
34 & 1359763168 & 22163717040384 & 155117520 & 2611304032624 & 35357670 & 611143639110 \\ 
36 & 5815457184 & 118021533366432 & 601080390 & 12582098181466 & 129644790 & 2785335811920 \\ 
38 & 24788842304 & 619313064407680 & 2333606220 & 60058408242252 & 477638700 & 12612104460780 \\ 
40 & 105340982248 & 3206924122635928 & 9075135300 & 284257070075212 & 1767263190 & 56773091159400 \\ 
\br
  &  \multicolumn{2}{c}{Stack polygons}
  &  \multicolumn{2}{c}{Ferrers diagrams}
  &  \multicolumn{2}{c}{Rectangular polygons} \\
\mr
$n$ & $p_n$ & $r_n$  & $p_n$ & $r_n$  & $p_n$ & $r_n$ \\
\mr
4 & 1 & 8 & 1 & 8 & 1 & 8 \\ 
6 & 2 & 66 & 2 & 66 & 2 & 66 \\ 
8 & 5 & 444 & 4 & 366 & 3 & 288 \\ 
10 & 13 & 2541 & 8 & 1640 & 4 & 900 \\ 
12 & 34 & 12840 & 16 & 6404 & 5 & 2280 \\ 
14 & 89 & 59113 & 32 & 22696 & 6 & 4998 \\ 
16 & 233 & 253600 & 64 & 74832 & 7 & 9856 \\ 
18 & 610 & 1029802 & 128 & 233312 & 8 & 17928 \\ 
20 & 1597 & 4002112 & 256 & 695680 & 9 & 30600 \\ 
22 & 4181 & 15005189 & 512 & 2000128 & 10 & 49610 \\ 
24 & 10946 & 54603436 & 1024 & 5578752 & 11 & 77088 \\ 
26 & 28657 & 193743969 & 2048 & 15166464 & 12 & 115596 \\ 
28 & 75025 & 672725072 & 4096 & 40336384 & 13 & 168168 \\ 
30 & 196418 & 2292470170 & 8192 & 105256960 & 14 & 238350 \\ 
32 & 514229 & 7685026612 & 16384 & 270135296 & 15 & 330240 \\ 
34 & 1346269 & 25392243845 & 32768 & 683188224 & 16 & 448528 \\ 
36 & 3524578 & 82826447752 & 65536 & 1705443328 & 17 & 598536 \\ 
38 & 9227465 & 267077278409 & 131072 & 4207935488 & 18 & 786258 \\ 
40 & 24157817 & 852322922488 & 262144 & 10274078720 & 19 & 1018400 \\ 
\br
\end{tabular}
\end{table}

\section{The exact generating functions \label{sec:sol}}

In this section we use the series for $r_n$ to find (numerically) the exact perimeter
generating functions for the radius of gyration of convex polygons.

The perimeter generating function for convex polygons was first obtained by
Delest and Viennot \cite{DV84a} using the method of algebraic languages and
later by several other authors using different methods \cite{GE88b,Lin88a,Kim88}:

\begin{equation}\label{eq:convex}
\PGf_{\rm Convex}(x)=\frac{x^2-6x^3+11x^4-4x^5}{(1-4x)^2}-\frac{4x^4}{(1-4x)^{3/2}}.
\end{equation}
\noindent
From this we see that the critical point $x_c=1/4$ (and thus $\mu=2$) while the
critical exponent $2-\alpha = -2$ (and thus $\alpha=4$), corresponding to the 
dominant double pole at $x=x_c$. In addition there is a sub-dominant square root 
correction. Informed by this result it is natural to assume that the generation function
for the mean-squared radius of gyration has a similar form. That is we assume
that $\RGf(x) = [A(x)+B(x)\sqrt{1-4x}]/(1-4x)^{\gamma}$, where $A(x)$ and $B(x)$ are
polynomials. Using the method of differential approximants \cite{AJG89a} we easily 
established that $\gamma =6$. Next we wrote a simple Maple routine to find such
a solution, that is we solve for the unknown coefficients $a_i$ and $b_i$ of $A(x)$ and $B(x)$.
We simply form the series expansion for $[A(x)+B(x)\sqrt{1-4x}]$, match the series
coefficients to those of $\RGf (x)(1-4x)^6$ and solve the resulting set of linear
equations in the coefficient $a_i$ and $b_i$. In this fashion we found a solution
with polynomials of degree 10 requiring no more than 22 unknown coefficients. Since
we have more than 50 known terms $r_{2n}$ there are at least 30 unused series coefficients
which serve as strong checks on the correctness of our solution. The generating
function for the mean-squared radius of gyration of convex polygons is:

\begin{eqnarray}\label{eq:RGconvex}
\fl
\RGf_{\rm Convex}(x)
 =\frac{2x^2(1-2x)(4-55x+388x^2-1058x^3+956x^4+2064x^5-6592x^6+6400x^7)}{(1-4x)^6} \nonumber \\ 
 -\frac{4x^4(15+22x-408x^2+1664x^3-3720x^4+3456x^5)}{(1-4x)^{11/2}}.
\end{eqnarray}
\noindent
From this we see that the critical exponent $\alpha+2\nu = 6$ and thus $\nu=1$. This should
be compared to the result for self-avoiding polygons $\nu=3/4$ \cite{Nienhuis82a}.
Physically, there is a simple argument for $\nu=1$. Convex polygons are 
relevant to the description of vesicles in the {\em inflated} regime, where they are 
space-filling, and since the radius of gyration measures a typical size of a polygon
$\ave{R^2}_n$ is proportional to a typical area and hence $\nu=1$ for convex polygons.
The value $\nu=3/4$ means that SAPs are much more ramified.

Directed and convex polygons was considered by Lin and Chang \cite{Lin88a}. They calculated
the full anisotropic generating function for directed and convex  polygons. In the isotropic
case which we consider here their result reduces to the very simple form

\begin{equation}\label{eq:dirconv}
\PGf_{\rm DirConv}(x)=\frac{x^2}{(1-4x)^{1/2}},
\end{equation}
\noindent
so we have $x_c=1/4$ while $2-\alpha=-1/2$ and thus $\alpha=5/2$.
As for the convex case we start by looking for a solution to $\RGf(x)$ of the same form,
that is $\RGf(x)=A(x)/(1-4x)^{\gamma}$, with $\gamma=9/2$ determined from differential
approximants. However we were not successful at first, so next we tried a solution of the 
same form as for convex polygons and found that

\begin{equation}\label{eq:RGdirconv}
\fl
\RGf_{\rm DirConv}(x)=\frac{-x^2+20x^3-48x^4+24x^5-168x^6+384x^7}{(1-4x)^{9/2}}
      +\frac{9x^2-44x^3+72x^4-32x^5}{(1-4x)^3}.
\end{equation}
\noindent
So in this case we find the critical exponent $\alpha+2\nu=9/2$ and thus as before $\nu=1$.

The model of staircase polygons is very well-known and much studied, dating back at least to
the work by P\'olya \cite{Polya69} who showed that 
$p_{2n}=\frac{1}{4n-2}{2n \choose n}$ for $n\geq 2$. This result was obtained
by Delest and Viennot \cite{DV84a} in the more elegant form 
$p_{2n+2}=C_n=\frac{1}{n+1}{2n \choose n}$, where $C_n$ are the famous and 
ubiquitous Catalan numbers. Consequently the generating function is

\begin{equation}\label{eq:stair}
\PGf_{\rm Stair}(x)=(1-2x-\sqrt{1-4x})/2,
\end{equation}
and $x_c=1/4$, while $2-\alpha=1/2$ and thus $\alpha=3/2$. As per the previous cases
we quite readily find the radius of gyration generation function

\begin{equation}\label{eq:RGstair}
\RGf_{\rm Stair}(x)=\frac{x(1-6x+24x^2-60x^3+64x^4)}{(1-4x)^{7/2}}-x,
\end{equation}
\noindent
and we see that $\alpha+2\nu=7/2$ and once again $\nu=1$.

Stack polygons were also considered by Lin and Chang \cite{Lin88a} and their result
for the generating function is

\begin{equation}\label{eq:stack}
\PGf_{\rm Stack}(x)=\frac{x^2(1-x)}{(1-3x+x^2)}.
\end{equation}
\noindent
The critical point is now given by the zero of $1-3x+x^2$ namely $x_c=0.381966011\ldots$ and
the critical exponent is $2-\alpha = -1$ or $\alpha=3$. In this case the radius of gyration
generation function is of the same form and again we have $\nu=1$. Explicitly we find that

\begin{equation}\label{eq:RGstack}
\fl
\RGf_{\rm Stack}(x)=
\frac{8x^2-54x^3+214x^4-489x^5+605x^6-386x^7+177x^8-120x^9+19x^{10}-x^{11}}{(1-3x+x^2)^5}.
\end{equation}

The generating function for Ferrers diagrams is trivial in that these polygons are
simply formed from a directed walk with $n-2$ right or up steps, extended at the
starting point with a horizontal step and at the end-point with a vertical step,
and then closed by straight lines to form a polygon with $2n$ steps. It immediately
follows that the generating function is

\begin{equation}\label{eq:ferrers}
\PGf_{\rm Ferrers}(x)=\frac{x^2}{(1-2x)},
\end{equation}
and we have $x_c=2$ and $\alpha=3$. The radius of gyration generation function is of the 
same form and with $\nu=1$,

\begin{equation}\label{eq:RGferrers}
\RGf_{\rm Ferrers}(x)=\frac{2x^2(4-7x+13x^2-10x^3+2x^4)}{(1-2x)^5}.
\end{equation}

Rectangular polygons are obviously the simplest case and the generating function
is simply

\begin{equation}\label{eq:rect}
\PGf_{\rm Rect}(x)=\frac{x^2}{(1-x)^2},
\end{equation}
so that $x_c=1$ and $\alpha=4$. The radius of gyration
generating function is found to be

\begin{equation}\label{eq:RGrect}
\RGf_{\rm Rect}(x)=\frac{2x^2(1+x)^2(4+x)}{(1-x)^6},
\end{equation}
and again we have $\nu=1$.

Now that these results for the radius of gyration of convex polygons are known
from the numerical work presented here it should be easier to prove them rigorously.

\section*{Acknowledgments}

We gratefully acknowledge financial support from the Australian Research Council.


\section*{References}

\begin{thebibliography}{10}

\bibitem{MSbook}
Madras N and Slade G 1993 {\em The self-avoiding walk\/} (Boston: Birkh\"auser)

\bibitem{HughesV1}
Hughes B~D 1995 {\em Random Walks and Random Environments, Vol {I} Random
  Walks\/} (Oxford: Clarendon)

\bibitem{RensburgBook}
Janse~van Renburg E~J 2000 {\em The statistical mechanics of interacting walks,
  polygons, animals and vesicles\/} (Oxford: Oxford University Press)

\bibitem{BM96a}
Bousquet-M\'elou M 1996 A method for the enumeration of various classes of
  column-convex polygons {\em Disc. Math.\/} {\bf 154} 1--25

\bibitem{Nienhuis82a}
Nienhuis B 1982 Exact critical point and critical exponents of {O}$(n)$ models
  in two dimensions {\em Phys. Rev. Lett.\/} {\bf 49} 1062--1065

\bibitem{GE88b}
Guttmann A~J and Enting I~G 1988 The number of convex polygons on the square
  and honeycomb lattices {\em J. Phys. A: Math. Gen.\/} {\bf 21} L467--474

\bibitem{IGE80e}
Enting I~G 1980 Generating functions for enumerating self-avoiding rings on the
  square lattice {\em J. Phys. A: Math. Gen.\/} {\bf 13} 3713--3722

\bibitem{JG99}
Jensen I and Guttmann A~J 1999 Self-avoiding polygons on the square lattice
  {\em J. Phys. A: Math. Gen.\/} {\bf 32} 4867--4876

\bibitem{IJ00a}
Jensen I 2000 Size and area of square lattice polygons {\em J. Phys. A: Math.
  Gen.\/} {\bf 33} 3533--3543

\bibitem{DV84a}
Delest M~P and Viennot G 1984 Algebraic languages and polyominoes enumeration
  {\em Theor. Comput. Scie.\/} {\bf 34} 169--206

\bibitem{Lin88a}
Lin K~Y and Chang S~J 1988 Rigorous results for the number of convex polygons
  on the square and honeycomb lattices {\em J. Phys. A: Math. Gen.\/} {\bf 21}
  2635--2642

\bibitem{Kim88}
Kim D 1988 The number of convex polyominoes with given perimeter {\em Discrete
  Math.\/} {\bf 70} 47--51

\bibitem{AJG89a}
Guttmann A~J 1989 Asymptotic analysis of power-series expansions in {\em Phase
  Transitions and Critical Phenomena\/} (eds. C~Domb and J~L Lebowitz) (New
  York: Academic) vol.~13 1--234

\bibitem{Polya69}
P\'olya G 1969 On the number of certain lattice polygons {\em J. Comb.
  Theory\/} {\bf 6} 102--105

\end{thebibliography}
\end{document}